\documentclass[10pt]{iopart}
\pdfoutput=1 %This is for the ArXiv submission only
\usepackage[pdftex]{graphicx,color}
\usepackage{amssymb}
\usepackage{epsfig}
\usepackage{cite}
\usepackage{float}
\usepackage{bm}% bold math
\expandafter\let\csname equation*\endcsname\relax
\expandafter\let\csname endequation*\endcsname\relax
\usepackage{amsmath}
\usepackage{amsmath,amssymb,eucal,graphicx,float,epstopdf}
\usepackage[breaklinks=true,colorlinks=true,linkcolor=blue,urlcolor=blue,citecolor=blue]{hyperref}
\DeclareSymbolFont{usualmathcal}{OMS}{cmsy}{m}{n}
\DeclareSymbolFontAlphabet{\mathcal}{usualmathcal}

\begin{document}
\title{Random sequential covering of a one-dimensional lattice by $k$-mers}
	
\author{Pascal Viot}
\address{Sorbonne  Universit\'{e}, Laboratoire de  Physique  Th\'{e}orique de la Mati\`{e}re Condens\'{e}e, Paris 75005, 
de la France Mati\`{e}re Condens\'{e}e (UMR CNRS 7600)}	
\author{P. L. Krapivsky}
\address{Department of Physics, Boston University, Boston, MA 02215, USA}
\address{Santa Fe Institute, Santa Fe, NM 87501, USA}

\begin{abstract} 
In random sequential covering, identical objects are deposited randomly, irreversibly, and sequentially; only attempts that increase coverage are accepted. The process continues indefinitely on an infinite substrate, and we analyze the dynamics of random sequential covering of $\mathbb{Z}$  using $k$-mers. We introduce a method that provides a comprehensive solution to the dynamics of this process. We derive explicit solutions for trimers, tetramers, and pentamers; we study numerically random sequential covering by longer polymers ($k>5$). 
\end{abstract}

\eads{\mailto{pascal.viot@sorbonne-universite.fr}, ~\mailto{pkrapivsky@gmail.com}}

\section{Introduction}

An old and venerable mathematical problem is to find the best covering of Euclidean spaces by spheres\cite{Conway}, or other identical objects\cite{Torquato10,Chen2014,Jiao2009,RamirezGonzalez2023}.  Space filling can be regular and random\cite{Flatto1977,Penrose23}. This a priori rather academic issue has unexpected connections with number theory\cite{Rogers,GL87,Kannan1987,Torquato10}, and applications to diverse subjects including shotgun sequencing in  genomics\cite{Ewens2005,Athreya04}, ballistics\cite{Hall88}, topological data analysis\cite{Bobrowski17,Tillmann22,Leykam2023}, stochastic optimization \cite{ZZ2008,Gomez2019}, immunology\cite{Moran}, wireless network protocols\cite{Lan,Baccelli2009,Baccelli2010}, ground states of interacting particles \cite{Cohn07,Cohn-Maryna22,Torquato18,Torquato10}, etc.

Previous studies have explored coverings of the ring by randomly placed arcs of identical lengths\cite{Stevens39} and random lengths $l_i$\cite{Shepp72a,Shepp72b,Siegel78,Siegel79,Siegel82}. Shepp\cite{Shepp72a,Shepp72b} obtained a perfect covering of the circle when the cumulative sum of $n^{-2}\sum_{1\leq i\leq n} e^{l_i}$ diverges. Multiple variants of this model have been studied\cite{Kahane,Coffman1994,Shepp87,Janson86,Jonasson08}, e.g., generalizations to Euclidean and non-Euclidean spaces \cite{GILBERT1965,Solomon1986}. The mean number of arcs covering a point on the circle can become infinite, indicating a lack of inefficiency. In contrast, the random covering process \cite{Krapivsky23} is a model where objects arrive sequentially, and the deposition attempts are accepted if they increase the overall coverage; the process stops upon achieving complete coverage.

The random sequential covering (RSC) of a one-dimensional lattice by `polymers' of length $k$ (shortly, $k$-mers) has been studied recently \cite{Krapivsky23}. A few analytical results have been established for arbitrary $k$. Only the simplest RSC by dimers ($k=2$) was solved, and the trimer model ($k=3$) already seemed intractable as equations for occupation probabilities looked hierarchical. Remarkably, for every $k$, there is a finite set of occupation probabilities of finite segments satisfying a closed set of equations. These equations are coupled but linear and, in principle, solvable for arbitrary $k$. We have succeeded in solving these equations using a pedestrian approach. Finding the general solution is an intriguing challenge. The straightforward calculations are feasible only for small $k$ as the number of equations roughly increases as $k^2$. Therefore, we primarily focus on the cases of trimers, tetramers, and pentamers ($k=3,4,5$).

The rest of this paper is organized as follows. In section \ref{sec:model}, we present the model and the method to obtain the time evolution of the process.  In sections \ref{sec:trimers}, \ref{sec:tetramers}, and \ref{sec:pentamers}, we apply the method to the trimer, tetramer, and pentamer models, respectively.

The complexity of the solution grows as $k^2$, forcing us to limit our analytical work to $k\leq 5$. In section \ref{sec:sim}, we numerically investigate the RSC processes with $k>5$. We also employ extrapolation techniques to estimate the continuum limit. In Sec.~\ref{sec:discussion}, we discuss our results and suggest potential applications to other models.

\section{The model}
\label{sec:model}

We consider the RSC processes on an infinite one-dimensional lattice. We always assume that the lattice is initially empty. Each $k$-mer occupies $k$ consecutive sites. The deposition events are random. An attempt is accepted if the $k$-mer covers at least one empty site; otherwise, the attempt is rejected. The RSC process stops when all lattice sites are covered. 

Denote by $\pi_j^{(k)}(t)$ the density of sites covered $j$ times. The definition of the RSC by $k-$mers implies that each site can be covered by at most $k$ polymers, so $\pi_j^{(k)}\equiv 0$  for $j>k$. The exact expressions for $\pi_j^{(k)}(t)$ with $j=0,1,\ldots,k$ constitute a rather complete description of the dynamics of the model. As with any `exactly solvable' model, it could be difficult to compute more complicated correlation functions. In the RSC models, the occupation probabilities $\pi_j^{(k)}(t)$ do not satisfy a closed system of equations, forcing us to consider more involved joint probabilities. We limit ourselves to computing $\pi_j^{(k)}(t)$ and required joint probabilities.

The conservation of the total probability gives the sum rule
\begin{equation}
	\sum_{i=0}^k \pi_j^{(k)}(t)=1
\end{equation}
To determine the occupation probabilities we rely on the joint probabilities $p_{i_1,i_2,...,i_l}$ of finding a sequence of $l$ consecutive sites with given `occupation' numbers $i_1,\ldots,i_l$. The left-right symmetry yields
\begin{equation}
\label{eq:sumrules}
	p_{i_1,i_2,...,i_l}=p_{i_l,i_{l-1},...,i_1}
\end{equation}
A deposition attempt is accepted when at least one of the $k$  sites is empty. This observation tells us that for any site with index $i$, the next-neighbor indices are always equal or differ by one. Below, we rely on these observations to deduce the evolution equations for the relevant occupation probabilities. In calculations, we also rely on the sum rules
\begin{equation}
	\sum_{i_l}p_{i_1,i_2,...,i_{l-1},i_l}=p_{i_1,i_2,...,i_{l-1}}
\end{equation} 

The evolution equations for the occupation probabilities depend on $p_{i_1,i_2,...,i_k}$ because these joint probabilities represent all possible deposition events. Yet the time evolution of some joint probabilities $p_{i_1,i_2,...,i_k}$ depends on higher-order joint probabilities. For random sequential adsorption (RSA) in one dimension\cite{Evans93,Talbot00}, the time evolution of intervals is independent, the so-called screening property. This feature allows one to obtain a closed set of equations. In most models, even in one dimension, where the hierarchy of equations is infinite, there is no hope of finding exact solutions.  For RSC by $k$-mers, however, the hierarchy of equations is finite, involving joint probabilities of sequences of $2k-1$ sites. This feature is the key to the solvability of the model. In contrast to the RSA by $k$-mers explicitly solvable for all $k$, the complexity of the computations (the number of equations that ought to be solved) for the RSC processes quickly increases with $k$, and in sections \ref{sec:trimers}--\ref{sec:pentamers} we explicitly solve the models with $k=3,4,5$. 

We remark that the $k\to\infty$ limit is especially intriguing as it can be re-interpreted as the continuous RSC, the RSC of the line $\mathbb{R}$ by sticks of unit length. The occupation probabilities $\pi_j$ with $j\geq 0$ are all non-trivial for the continuous RSC. These occupation probabilities are unknown apart from the fraction $\pi_0$ of the uncovered line\cite{Krapivsky23}. In contrast, the continuous RSA on the line is a famous problem solved by R\'{e}nyi \cite{Renyi58}. 

As a warm up, we show how to compute some simple joint probabilities. The densities of empty strings ($\circ$ denotes an empty site)
\begin{equation}
	\label{Em}
	p_{\underbrace{\circ \cdots\circ}_m }\equiv E_m^{(k)} = \text{Prob}[\underbrace{\circ \cdots\circ}_m]
\end{equation}
satisfy 
\begin{equation}
	\label{Emt-3}
	\tfrac{dE_m^{(k)}}{dt} =-(m+k-1)E_m^{(k)}, \qquad m\geq 1
\end{equation}
from which
\begin{equation}
	\label{Em-3-sol}
	E_m^{(k)}=e^{-(m+k-1 )t}
\end{equation}
Specializing \eqref{Em-3-sol} to $m=1$ yields the density of empty sites 
\begin{equation}
	\label{empty-k}
	\pi_0^{(k)}  = e^{-k t}
\end{equation}

Another simple family ($\bullet$ denotes a single-covered site)
\begin{equation}
	\label{Vm}
	p_{\bullet\underbrace{\circ \cdots\circ}_m\bullet }\equiv 	V_m^{(k)} = \text{Prob}[\bullet\underbrace{\circ \cdots\circ}_m \bullet]
\end{equation}
describes the voids. The definition of the RSC process implies that an occupied site adjacent to an empty site is single-covered. Expressing the density of voids through the density of empty strings (see, e.g., \cite{KRB}) we obtain
\begin{equation}
	\label{V-E}
	V_m^{(k)} = E_m^{(k)} - 2E_{m+1}^{(k)} + E_{m+2}^{(k)}
\end{equation}
Combining \eqref{V-E} with \eqref{Em-3-sol} we get 
\begin{equation}
	\label{Vm-sol}
	V_m^{(k)} = e^{-(m+k-1)t}\,(1-e^{-t})^2
\end{equation}

The third family
\begin{equation}
	\label{Sm}
	S_m^{(k)} = \text{Prob}[\underbrace{\circ \cdots\circ}_m \bullet] = \text{Prob}[\bullet\underbrace{\circ\cdots\circ}_m]
\end{equation}
is also expressible through the densities of empty strings, $S_m^{(k)} = E_m^{(k)} -E_{m+1}^{(k)}$, and hence
\begin{equation}
	\label{Sm-sol}
	S_m^{(k)} =e^{-(m+k-1)t}\,(1-e^{-t})
\end{equation}

The weighted sum 
\begin{equation}
\label{sum-k}
\textsf{M}_k=\sum_{j=2}^k (j-1)\pi_j^{(k)}(t) 
\end{equation}
is a combination of a constant and two exponents \cite{Krapivsky23}
\begin{equation}
\label{sum-k-sol}
\textsf{M}_k=\frac{k-1}{k+1}-(k-1)e^{-kt}+\frac{k(k-1)}{k+1}\,e^{-(k+1)t}
\end{equation}
The normalization requirement gives
\begin{equation}
\label{norm-k}
\sum_{j=0}^k \pi_j^{(k)}(t) = 1
\end{equation}
Combining \eqref{sum-k} and \eqref{norm-k}, one obtains the average coverage $\overline{n(t)}$
\begin{equation}\label{eq:meancov}
\overline{n(t)}=\sum_{l=1}^{k} j\pi_j^{(k)}(t)=\frac{2k}{k+1}-k e^{-kt}+\frac{k(k-1)}{k+1}\,e^{-(k+1)t}
\end{equation}
The average coverage at saturation, $\overline{n(\infty)}=\frac{2k}{k+1}$, is a slowly increasing function of $k$ approaching to the double coverage when $k\to\infty$. One then expects a fast decay of $\pi_j^{(k)}(\infty)$ when $j$ increases.

Note that the short-time expansion of the densities can be obtained easily because they correspond to the deposition of first $k$-mers or building a single cluster.
Therefore,  enumerating the configurations, the  densities are given by
\begin{equation}
 \pi_j^{(k)} = \binom{k}{j} \frac{2^{j-1}}{j!}\, t^j +0(t^{j+1})
\label{eq:shorttime}
\end{equation}

Specializing Eqs.~\eqref{empty-k}, \eqref{sum-k-sol} and \eqref{norm-k} to dimers ($k=2$), the complete exact solution can be obtained immediately
\begin{equation}\label{jammed12}
	\begin{split}
\pi_1^{(2)}&=\frac{2}{3}-\frac{2}{3}e^{-3t}\\
\pi_2^{(2)}&=\frac{1}{3}-e^{-2t}+\frac{2}{3}e^{-3t}
	\end{split}
\end{equation}

At short time $T\sim t$ and Eq.\eqref{jammed12} confirms Eq.\eqref{eq:shorttime}.
Hence for the RSC by dimers, the single-covered occupation probability at saturation is twice greater than the double-covered occupation probability\cite{Krapivsky23}.

We now separately consider the RSC by trimers, tetramers, and pentamers. To avoid cluttering of formulas, we shortly write $\pi_j$ instead of $\pi_j^{(k)}$.

\section{Trimers}
\label{sec:trimers}

For the trimer model ($k=3$), we must compute $\pi_i$ with $i=0, 1, 2, 3$. Specializing \eqref{empty-k}--\eqref{norm-k} to trimers, the sum rules provides the following equations:
\begin{subequations}
\begin{align}
\label{p0-trimers}
&\pi_0 = e^{-3t}\\
\label{p23-trimers}
&\pi_2 + 2\pi_3  =  \tfrac{1}{2} - 2 e^{-3t} + \tfrac{3}{2} e^{-4t} \\
\label{norm-trimers}
&\pi_1+\pi_2+\pi_3 = 1-e^{-3t}
\end{align}
\end{subequations}

In contrast to the RSC by dimers, we cannot extract three densities $\pi_1, \pi_2, \pi_3$ from the two exact sum rules \eqref{p23-trimers}--\eqref{norm-trimers}, we need one more relation.  

We now introduce the joint probabilities $p_{ijk}$ for obtaining a complete solution of the kinetics. The evolution equations for the densities $\pi_0, \pi_1, \pi_2, \pi_3$ depend on the densities of strings of three adjacent sites
\begin{subequations}
\label{p0123}
\begin{align}
\label{p0-eq}
&\frac{d\pi_0}{dt} = -3p_{000}-4p_{001}-2p_{011}-p_{101}-2p_{012}\\
\label{p1-eq}
&\frac{d\pi_1}{dt} = 3p_{000} + 2p_{001}-2p_{011}-p_{101}  \\
\label{p2-eq}
&\frac{d\pi_2}{dt} = 2p_{001}+4p_{011} + 2p_{101}  \\
\label{p3-eq}
&\frac{d\pi_3}{dt} = 2p_{012}
\end{align}
\end{subequations}
These equations follow from the definition of the RSC by trimers. The notations for the densities of strings of three adjacent sites are self-explanatory. In Eqs.~\eqref{p0123}, we also used the left-right symmetry: $p_{001}=p_{100}$ [see \eqref{Sm}],  $p_{011}=p_{110}$, $p_{012}=p_{210}$, etc. 

We know some densities appearing in Eqs.~\eqref{p0123}:
\begin{subequations}
\label{p-short}
\begin{align}
\label{p000}
&p_{000}=\text{Prob}[\circ\circ\circ]=E_3=e^{-5t}\\
\label{p001}
&p_{001}=\text{Prob}[\circ\circ\bullet]=S_2=e^{-4t}\,(1-e^{-t})\\
\label{p101}
&p_{101}=\text{Prob}[\bullet\circ\bullet]=V_1=e^{-3t}\,(1-e^{-t})^2
\end{align}
\end{subequations}
Finding the densities $p_{011}$ and $p_{012}$ require a bit more work. One writes the evolution equations 
\begin{subequations}
\begin{align}
\label{p011-eq}
&\frac{dp_{011}}{dt} = -3p_{011}-p_{01110}+p_{000}\\
\label{p012-eq}
&\frac{d p_{012}}{dt} = -3p_{012}+p_{01110}+p_{001} 
\end{align}
\end{subequations}
These equations directly follow from the definition of the RSC and in addition to 3-sites joint probabilities involve a single the 5-sites joint probability. The corresponding density $p_{01110}$ satisfies 
\begin{align}
\label{p01110-eq}
\frac{dp_{01110}}{dt} = -6p_{01110}+p_{00000}
\end{align}
Using $p_{00000}=E_5=e^{-7t}$ we solve \eqref{p01110-eq} and find
\begin{align}
\label{p01110}
p_{01110} = e^{-6t}(1-e^{-t})
\end{align}
We now insert \eqref{p000}, \eqref{p001} and \eqref{p01110} into  (\ref{p011-eq}) and -(\ref{p012-eq})  and solve resulting equations to yield
\begin{subequations}
\label{p-long}
\begin{align}
\label{p011}
&p_{011}=\frac{e^{-3t}(1-e^{-t})\left(5+5e^{-t}-e^{-2t}+3e^{-3t}\right)}{12}\\
\label{p012}
&p_{012}=\frac{e^{-3t}(1-e^{-t})^2\left(7+2e^{-t}+e^{-2t}\right)}{12}
\end{align}
\end{subequations}

Plugging \eqref{p-short} and \eqref{p-long} into \eqref{p1-eq}--\eqref{p3-eq} and integrating we obtain the complete solution

\begin{equation}
	\begin{split}
		&\pi_1 =
		T\left(3-6 T+\frac{17}{3} T^{2}-\frac{17}{6} T^{3}+\frac{31}{30} T^{4}-\frac{7}{18} T^{5}+\frac{1}{14} T^{6}\right)  \\
		&\pi_2= T^2\left( 3-\frac{16}{3} T+\frac{25}{6} T^{2}-\frac{31}{15} T^{3}+\frac{7}{9} T^{4}-\frac{1}{7} T^{5} \right)   \\
		&\pi_3 =T^3 \left( \frac{2}{3}-\frac{4}{3} T+\frac{31}{30} T^{2}-\frac{7}{18} T^{3}+\frac{1}{14} T^{4} \right)
	\end{split}
	\label{sol:trimer}
\end{equation}
where $T$ is the "compressed" time defined by 
\begin{equation}\label{eq:compressed}
	T=1-e^{-t}
\end{equation}

%
%\begin{equation}
%\begin{split}
%&\pi_1 = \tfrac{173}{315}+\tfrac{11}{18}e^{-3t}-e^{-4t}-\tfrac{1}{5}e^{-5t}+\tfrac{1}{9}e^{-6t}-\tfrac{1}{14}e^{-7t}  \\
%&\pi_2=  \tfrac{253}{630}-\tfrac{11}{9}e^{-3t}+\tfrac{1}{2}e^{-4t}+\tfrac{2}{5}e^{-5t}-\tfrac{2}{9}e^{-6t}+\tfrac{1}{7}e^{-7t} \\
%&\pi_3 = \tfrac{31}{630}-\tfrac{7}{18}e^{-3t}+\tfrac{1}{2}e^{-4t}-\tfrac{1}{5}e^{-5t}+\tfrac{1}{9}e^{-6t}-\tfrac{1}{14}e^{-7t} 
%\end{split}
%\label{sol:trimer}
%\end{equation}
It is easy to check that the solution \eqref{sol:trimer} satisfies the sum rule \eqref{eq:meancov}. The short-time expansion of the densities are easy to read from Eq.\eqref{sol:trimer} and gives $\pi_1 \sim 3t,
\pi_2 \sim 3t^2,  \pi_3 \sim \frac{2}{3}t^3$, as expected (see Eq.\eqref{eq:shorttime}).
\begin{figure}[t]
	\centering
	\includegraphics[width=0.88\columnwidth]{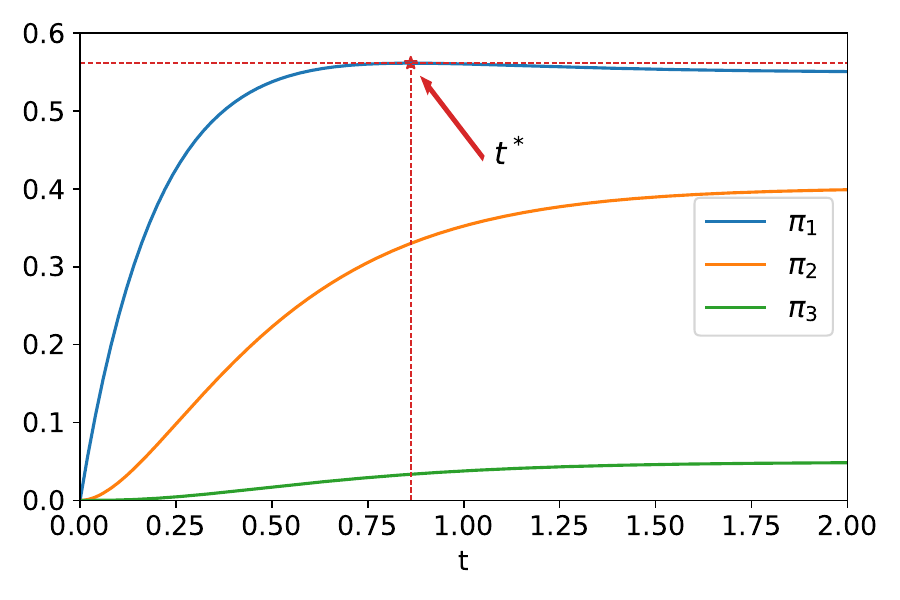}
\caption{Trimer model: densities $\pi_1(t), \pi_2(t)$ and $\pi_3(t)$ versus $t$. The fraction $\pi_1$ of single-occupied sites reaches the maximum at a finite time $t^*$ (vertical and horizontal lines are guides for an eye). } 
\label{fig:trimer}
\end{figure}

When $T=1$, one obtains the densities in the final (jammed) state 
\begin{equation}
\label{jammed:123}
\pi_1 = \frac{173}{315}\,, \quad \pi_2=  \frac{253}{630}\,, \quad \pi_3 = \frac{31}{630}
\end{equation}

Figure \eqref{fig:trimer} shows the time evolution of the densities: whereas $\pi_2$ and $\pi_3$ increased monotonously with time, $\pi_1$ displays a maximum at a finite time $t^*$. The maximal time is a root of a quartic equation; numerically $t^*=0.861987$. This maximum of the first layer density at a finite time is the consequence of the formation of the second and third layers that appear at late times, compared to the first-layer deposition. These behaviors differ from the dimer kinetics \cite{Krapivsky23} where both $\pi_1$ and $\pi_2$ are monotonously increasing functions of time.

\section{Tetramers}
\label{sec:tetramers}

For the tetramer model ($k=4$), we must compute $\pi_i$ with $i=0, \ldots, 4$. We already know $\pi_0=e^{-4t}$. The remaining densities satisfy two sum rules
\begin{subequations}
\begin{align}
\label{p234-tetramers}
&\pi_2 + 2\pi_3  + 3 \pi_4 =  \tfrac{3}{5} - 3 e^{-4t} + \tfrac{12}{5} e^{-5t} \\
\label{norm-tetramers}
&\pi_1 + \pi_2 + \pi_3 + \pi_4 = 1-e^{-4t}
\end{align}
\end{subequations}

 The density of maximally covered sites satisfies
\begin{equation}
\label{p4-eq}
\frac{d\pi_4}{dt} = 2 p_{0123}
\end{equation}

The governing equations for the remaining densities $\pi_1, \pi_2, \pi_3$ can be written in the matrix form
\begin{equation}
\label{eq:mers}
\frac{d\mathbf{\Pi}}{dt} = \mathbb{A} \mathbf{P}
\end{equation}
with vectors 
\begin{equation*}
\mathbf{\Pi} = 
\begin{pmatrix}\pi_1 \\ \pi_2 \\ \pi_3 
\end{pmatrix}, \qquad
\mathbf{P} = 
\begin{pmatrix}p_{0000} \\ p_{0001} \\ p_{0011} \\ p_{0012}  \\ p_{0111} \\ p_{0112} \\ p_{0122}  \\ p_{1011}\\ p_{1012}\\ p_{1001}
\end{pmatrix}
\end{equation*}
and matrix 
\begin{equation*}
\mathbb{A} = \begin{pmatrix} 4 & 4 & 0 & 2 &-4 &-2 & 0  &-4 &-2 & 0 \\ 
                                0 & 2 & 4 & 0 & 6  & 2 &-2 & 6 & 2  & 2\\
                                0 & 0 & 0 & 2 & 0  & 2 & 4 & 0 & 2 & 0  \end{pmatrix}
\end{equation*}

The joint probabilities constituting $\mathbf{P}$ satisfy
\begin{equation}
\label{P-4}
	\begin{split}
		\frac{d p_{0000}}{dt}&=-7p_{0000}\\
		\frac{d p_{0001}}{dt}&=-6p_{0001}+p_{0000}\\
		\frac{d p_{0011}}{dt}&=-5p_{0011}-	p_{0011110}+p_{0000}\\
		\frac{d p_{0012}}{dt}&=-5p_{0012}+	p_{0011110}+p_{0001}\\
		\frac{d p_{0111}}{dt}&= -4p_{0111} -	2 p_{011110}+p_{0000}\\
		\frac{d p_{0112}}{dt}&= -4p_{0112}+ p_{011110}+p_{0001}\\
		\frac{d p_{0122}}{dt}&= -4p_{0122}+ p_{011110}+p_{0011}\\
		\frac{d p_{0123}}{dt}&= -4p_{0123}+ p_{0122210}+p_{0012}\\
		\frac{d p_{1011}}{dt}&= -4p_{1011}- p_{1011110} +p_{0011}+ p_{1000}\\
		\frac{d p_{1012}}{dt}&= -4p_{1012}+p_{1011110} +p_{1001}+p_{0012}\\
		\frac{d	p_{1001}}{dt}&=-5p_{1001}+2p_{0001}\\
	\end{split}
\end{equation}

In the case of the RSC by tetramers 
\begin{equation*}
\{E_m,V_m,S_m\}=e^{-(m+3)t}\,\{1, (1-e^{-t})^2, (1-e^{-t})\}
\end{equation*}
Hence, we fix three densities appearing in Eqs.~\eqref{P-4}: 
\begin{subequations}
\label{p4-short}
\begin{align}
\label{p0000}
&p_{0000}=E_4=e^{-7t}\\
\label{p0001}
&p_{0001}=S_3=e^{-6t}\,(1-e^{-t})\\
\label{p1001}
&p_{1001}=V_2=e^{-5t}\,(1-e^{-t})^2
\end{align}
\end{subequations}

Higher-order joint probabilities also appear in Eqs.~\eqref{P-4}. We define three families
\begin{subequations}
\begin{align}
	\label{Tm}
	T_{\mu} &= \text{Prob}[\underbrace{\circ \cdots\circ}_{m_1}
	\underbrace{\bullet \cdots\bullet}_n \underbrace{\circ \cdots\circ}_{m_2} ]   \\
	\label{Um}
	U_{\mu+1} &= \text{Prob}[\bullet\underbrace{\circ \cdots\circ}_{m_1}
	\underbrace{\bullet \cdots\bullet}_n \underbrace{\circ \cdots\circ}_{m_2} ] \\
	\label{Wm}
	W_{\mu+1} &= \text{Prob}[\underbrace{\circ \cdots\circ}_{m_1}\bullet
	\underbrace{\blacktriangle \cdots\blacktriangle}_{n-1}\bullet\underbrace{\circ \cdots\circ}_{m_2} ] 
\end{align}
\end{subequations}
where we shortly write $\mu=m_1+n+m_2$ and denote by $\blacktriangle$ a double-covered site. It is easy to show that 
\begin{equation}
\label{TUV}
	\{T_{\mu},U_{\mu+1},W_{\mu+1}\}=e^{-(m_1+m_2+2n-2)t}\,\{(1-e^{-t}),(1-e^{-t})^2, (1-e^{-t})^2\}
\end{equation}
Using \eqref{TUV} we fix higher-order joint probabilities appearing in Eqs.~\eqref{P-4}:
\begin{equation}
\label{p4-long}
\begin{split}
p_{011110} & = e^{-8t}(1-e^{-t})\\
p_{0011110} & = e^{-9t}(1-e^{-t})\\
p_{0122210} & = e^{-8t}(1-e^{-t})^2\\
p_{1011110} & = e^{-8t}(1-e^{-t})^2
\end{split}
\end{equation}
Using \eqref{p4-short} and \eqref{p4-long} we solve Eqs.~\eqref{P-4} and find remaining joint probabilities. We then solve Eqs.~\eqref{eq:mers} and arrive at

\begin{equation}
	\begin{split}
		\pi_1&=T\left(4-12 T+\frac{56}{3} T^{2}-\frac{109}{6} T^{3}+\frac{68}{5} T^{4}-\frac{134}{15} T^{5}+\frac{472}{105} T^{6}-\frac{27}{20} T^{7}+\frac{8}{45} T^{8}\right)\\
		\pi_2&= T^2\left( 6-\frac{52}{3} T+\frac{77}{3} T^{2}-26 T^{3}+\frac{1187}{60} T^{4}-\frac{451}{42} T^{5}+\frac{293}{80} T^{6}-\frac{59}{90} T^{7}+\frac{1}{25} T^{8}
		\right)\\
		\pi_3&=	 T^3 \left( \frac{8}{3}-\frac{53}{6} T+\frac{68}{5} T^{2}-\frac{383}{30} T^{3}+\frac{839}{105} T^{4}-\frac{131}{40} T^{5}+\frac{7}{9} T^{6}-\frac{2}{25} T^{7}\right) 
		\\
		\pi_4&=	 T^4 \left( \frac{1}{3}-\frac{6}{5} T+\frac{23}{12} T^{2}-\frac{367}{210} T^{3}+\frac{77}{80} T^{4}-\frac{3}{10} T^{5}+\frac{1}{25} T^{6}\right) 
	\end{split}
	\label{eq:tetramersol}	
\end{equation}
where $T=1-e^{-t}$.

The short-time expansion of the densities are  easy to read from Eq.\eqref{sol:trimer} and gives $\pi_1 \sim 4t,
\pi_2 \sim 6t^2,  \pi_3 \sim \frac{8}{3}t^3$, and $\pi_4\sim \frac{t^4}{3}$ as expected (see Eq.\eqref{eq:shorttime}).
As a consistency check, one can verify that  Eqs.~\eqref{eq:tetramersol} satisfies the sum rule \eqref{eq:meancov}. The densities in the final (jammed) state are 

\begin{equation}
\label{jammed:1234}
\pi_1 = \frac{617}{1260}\,, \quad \pi_2=  \frac{10723}{25200}\,, \quad \pi_3 = \frac{1007}{1260}\,, \quad \pi_4 = \frac{41}{8400}
\end{equation}
There is a strong empirical evidence that $\pi_1^{(k)}$ decreases with $k$, while $\pi_j^{(k)}$ with $j\geq 2$ are increasing functions of $k$.

\begin{figure}[t]
	\centering
	\includegraphics[width=0.8\columnwidth]{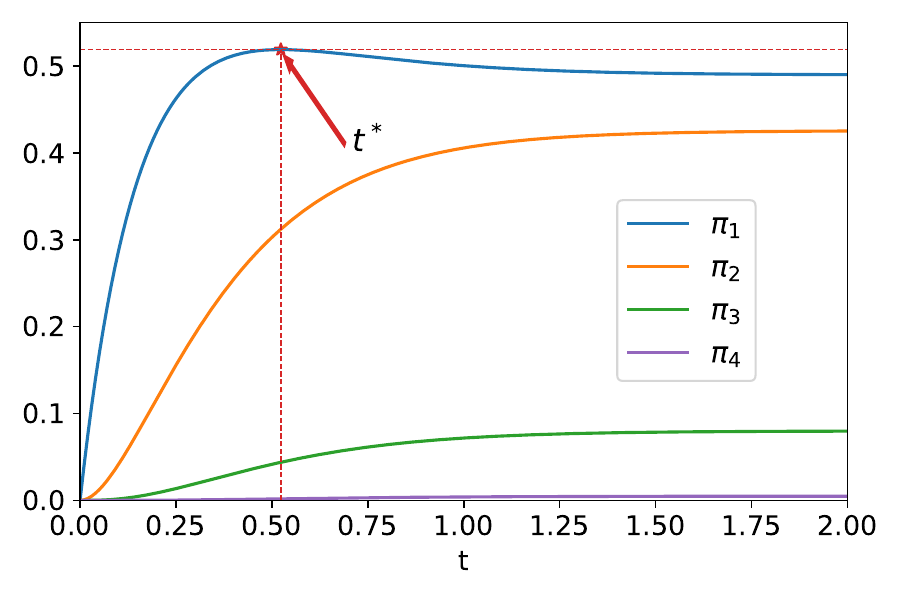}
\caption{Tetramer model: densities $\pi_1(t), ~\pi_2(t), ~\pi_3(t), ~\pi_4(t)$ versus $t$. The fraction $\pi_1$ of single-occupied sites reaches the maximum at a finite time $t^*$ (vertical and horizontal lines are guides for an eye).} 
	\label{fig:tetramer}
\end{figure}

Figure \eqref{fig:tetramer} illustrates the time evolution of the densities:  $\pi_2$,  $\pi_3$, and $\pi_4$ are increasing functions over time, while the density $\pi_1$  exhibits a maximum at a finite time. This time is a root of a fifth-order equation; numerically, $t^*=0.5232966529$. Compared to the trimer model, the maximum of $\pi_1(t)$  occurs more rapidly, and the observed overshoot is more pronounced. To provide a narrative overview of the line coverage, we can say that the process begins with the formation of the first layer, during which one site in every two becomes covered, while one-third of the sites are occupied twice.

Thus, the method involves examining the past, which ultimately requires higher-order joint probabilities of length $2k-1$ (where one or two $k$-mers have already been deposited). This observation indicates that the range of correlation is limited. Additionally, it explains why the expressions for occupation probabilities consist of sums of exponential terms ranging from $e^{-kt}$ to $e^{-(3k-2)t}$ with amplitudes being rational numbers. Therefore, the occupation probabilities at jamming are rational numbers for every finite $k$. For the continuous RSC, equivalently in the $k\to\infty$ limit, the jammed occupation probabilities $\pi_j(\infty)$ with $j\geq 1$ could be non-rational.

\section{Pentamers}  
\label{sec:pentamers}
For the pentamer model ($k=5$), we must compute $\pi_i$ with $i=0, \ldots, 5$. We already know $\pi_0=e^{-5t}$. The remaining densities satisfy two sum rules
\begin{subequations}
	\begin{align}
		\label{p2345-pentamers}
		&\pi_2 + 2\pi_3  + 3 \pi_4 +4 \pi_5=  \frac{2}{3}-4 {\mathrm e}^{-5 t}+\frac{10 {\mathrm e}^{-6 t}}{3}\\
		\label{norm-pentamers}
		&\pi_1 + \pi_2 + \pi_3 + \pi_4 + \pi_5 = 1-e^{-5t}
	\end{align}
\end{subequations}
As demonstrated for the trimer and tetramer models, the system is solvable analytically because the hierarchy involves joint probabilities of finite higher orders. While the trimer and tetramer models require the computation of $5$ and $11$ joint probabilities, the pentamer model presents a greater challenge as the number of joint probabilities increases to $26$. The RSC of pentamers requires to consider occupation probabilities $\pi_0,\pi_1,\pi_2$, $\pi_3$, $\pi_4$ and $\pi_5$. We know $\pi_0(t)=e^{-5t}$. The kinetic equations for these occupation probabilities can be written again in a matrix form \eqref{eq:mers} with 
\begin{equation}
\label{Pi-5}
\mathbf{\Pi}^T = (\pi_1, \pi_2, \pi_3, \pi_4, \pi_5)
\end{equation}
and
\begin{equation}
\label{AP-5}
\begin{split}
 \mathbb{A}^T=
\left(\begin{array}{ccccc}
	 5 & 0 & 0 & 0 & 0 
	\\
	 6 & 2 & 0 & 0 & 0 
	\\
	 2 & 4 & 0 & 0 & 0 
	\\
	 4 & 0 & 2 & 0 & 0 
	\\
	 -2 & 6 & 0 & 0 & 0 
	\\
	 0 & 2 & 2 & 0 & 0 
	\\
	 2 & -2 & 4 & 0 & 0 
	\\
	 2 & 0 & 0 & 2 & 0 
	\\
	 -6 & 8 & 0 & 0 & 0 
	\\
	 -4 & 4 & 2 & 0 & 0 
	\\
	 -2 & 0 & 4 & 0 & 0 
	\\
	 -2 & 2 & 0 & 2 & 0 
	\\
	 0 & -4 & 6 & 0 & 0 
	\\
	 0 & -2 & 2 & 2 & 0 
	\\
	 0 & 0 & -2 & 4 & 0 
	\\
	 0 & 0 & 0 & 0 & 2 
	\\
	 1 & 2 & 0 & 0 & 0 
	\\
	 -2&6 & 0 & 0 & 0 
	\\
	 0 & 2 & 2 & 0 & 0 
	\\
	-6& 8 & 0 & 0 & 0 
	\\
	 -4 & 4 & 2 & 0 & 0 
	\\
	-2& 0 & 4 & 0 & 0 
	\\
	 -2 & 2 & 0 & 2 & 0 
	\\
	 -3 & 4 & 0 & 0 & 0 
	\\
	 -4 & 4 & 2 & 0 & 0 
	\\
	 -1 & 0 & 2 & 0 & 0 
\end{array}\right) & \qquad
	\mathbf{P}=\left(\begin{array}{c}
		p_{00000} 
		\\
		p_{00001} 
		\\
		p_{00011} 
		\\
		p_{00012} 
		\\
		p_{00111} 
		\\
		p_{00112} 
		\\
		p_{00122} 
		\\
		p_{00123} 
		\\
		p_{01111} 
		\\
		p_{01112} 
		\\
		p_{01122} 
		\\
		p_{01123} 
		\\
		p_{01222} 
		\\
		p_{01223} 
		\\
		p_{01233} 
		\\
		p_{01234} 
		\\
		p_{10001} 
		\\
		p_{10011} 
		\\
		p_{10012} 
		\\
		p_{10111} 
		\\
		p_{10112} 
		\\
		p_{10122} 
		\\
		p_{10123} 
		\\
		p_{11011} 
		\\
		p_{11012} 
		\\
		p_{21012} 
	\end{array}\right)	
\end{split}
\end{equation}

The probabilities $p_{ijklm}$ represent sequences of five consecutive sites with occupation numbers $ijklm$. Similarly to the trimer and tetramer problems, the left-right symmetry yields $p_{ijklm}= p_{mlkji}$.

In the case of the RSC by pentamers
\begin{equation*}
	\{E_m,V_m,S_m\}=e^{-(m+4)t}\,\{1, (1-e^{-t})^2, (1-e^{-t})\}
\end{equation*}
from which
\begin{subequations}
	\label{p5-short}
	\begin{align}
		\label{p00000}
		&p_{00000}=E_5=e^{-9t}\\
		\label{p00001}
		&p_{0001}=S_4=e^{-8t}\,(1-e^{-t})\\
		\label{p10001}
		&p_{1001}=V_3=e^{-8t}\,(1-e^{-t})^2
	\end{align}
\end{subequations}

Details of the derivation of other joint probabilities are relegated to \ref{sec:pentamerappendix}. Collecting all results, one solves Eqs.~\eqref{eq:mers} to give
\begin{equation}
\begin{split}
\pi_1=&T\left( 5-20 T+\tfrac{130}{3} T^{2}-\tfrac{125}{2} T^{3}+\tfrac{711}{10} T^{4}-\tfrac{209}{3} T^{5}+\tfrac{1171}{21} T^{6}-\tfrac{98}{3} T^{7}+\tfrac{229}{18} T^{8}-\tfrac{44}{15} T^{9}+\tfrac{10}{33} T^{10} \right)\\
\pi_2=&T^2\left(10-40 T+\tfrac{260}{3} T^{2}-\tfrac{662}{5} T^{3}+\tfrac{13949}{90} T^{4}-\tfrac{29087}{210} T^{5}+\tfrac{25489}{280} T^{6}-\tfrac{318839}{7560} T^{7}+\tfrac{27589}{2100} T^{8}\-\tfrac{3973}{1540}T^{9}+\tfrac{733}{2520} T^{10}-\tfrac{5}{312} T^{11} \right)\\
\pi_3=&T^3\left(\tfrac{20}{3}-\tfrac{185}{6} T+\tfrac{2119}{30} T^{2}-\tfrac{4726}{45} T^{3}+\tfrac{23411}{210} T^{4}-\tfrac{9097}{105} T^{5}+\tfrac{367817}{7560} T^{6}-\tfrac{39997}{2100} T^{7}+\tfrac{22597}{4620} T^{8}\right)\\
\pi_4=&T^4\left(\tfrac{5}{3}-\tfrac{127}{15} T+\tfrac{1837}{90} T^{2}-\tfrac{6401}{210} T^{3}+\tfrac{8637}{280} T^{4}-\tfrac{163837}{7560} T^{5}+\tfrac{7289}{700} T^{6}-\tfrac{1367}{420} T^{7}\right.\\&\left. +\tfrac{1499}{2520} T^{8}-\tfrac{5}{104} T^{9} \right)\\
\pi_5=&T^5\left(\tfrac{2}{15}-\tfrac{32}{45} T+\tfrac{367}{210} T^{2}-\tfrac{1081}{420} T^{3}+\tfrac{18679}{7560} T^{4}-\tfrac{3299}{2100} T^{5}+\tfrac{269}{420} T^{6}-\tfrac{383}{2520} T^{7}+\tfrac{5}{312} T^{8} \right)
\end{split}
\end{equation}

\begin{figure}[t]
	\centering
	\includegraphics[width=0.8\columnwidth]{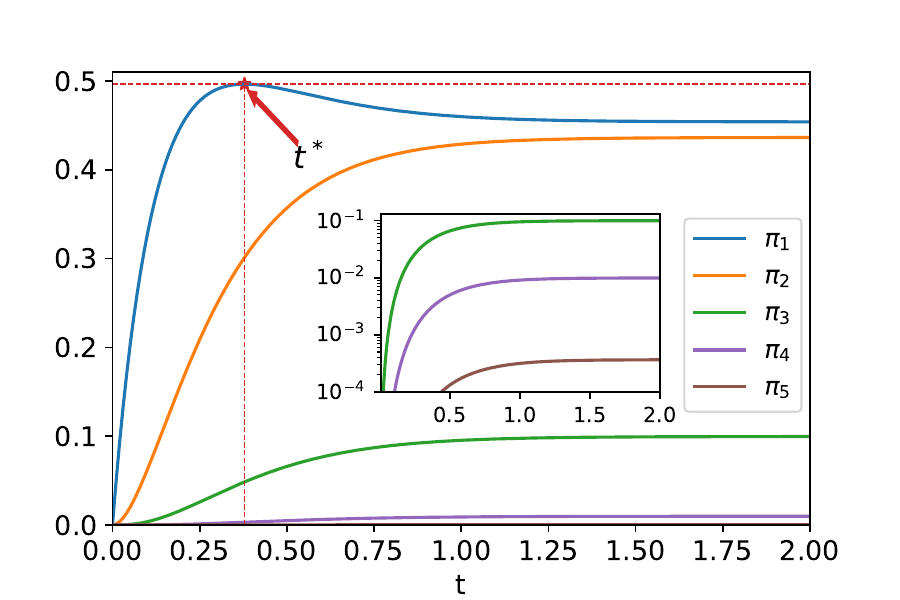}
\caption{Pentamer model: densities $\pi_1(t), \pi_2(t), \pi_3(t), \pi_4(t), \pi_5(t)$ versus $t$. The fraction $\pi_1$ of single-occupied sites reaches the maximum at a finite time $t^*$ (vertical and horizontal lines are guides for an eye). The inset highlights $\pi_3(t)$, $\pi_4(t)$ and $\pi_5(t)$ which remain very small for all times; the density $\pi_5(t)$ is invisiable in the main plot. } 
	\label{fig:pentamer}
\end{figure}
The densities in the final (jammed) state are
\begin{equation}
	\label{jammed:12345}
	\begin{split}
	&\pi_1 = \tfrac{629}{1386}\simeq 0.4538239538\\
	&\pi_2 = \tfrac{294797}{675675}\simeq 0.4362999963 \\
	& \pi_3 =\tfrac{76931}{772200}\simeq 0.09962574463\\
	& \pi_4 = \tfrac{347}{35100}\simeq 0.009886039886\\
	& \pi_5 = \tfrac{179}{491400}\simeq 0.0003642653643
	\end{split} 
\end{equation}
This solution reinforces the general trend that the single occupation probability $\pi_1$ decreases as $k$ increases,while all multiple occupation probabilities increase with $k$. 

Figure \eqref{fig:pentamer} illustrates the time evolution of the densities:  $\pi_1 $, $\pi_2$, $\pi_3$,$\pi_4$   and $\pi_5$ are increasing functions over time, but  $\pi_1$ also exhibits a maximum at a finite time, which occurs sooner than in the tetramer model.

The fraction $\pi_1$ of single-occupied sites reaches the maximum at a finite time that is a root of a polynomial of the sixth-order; numerically $t^*=0.3784190341$.  As expected, the maximum of $\pi_1(t)$ occurs more rapidly, and the observed overshoot is more pronounced than in the trimer and tetramer models. Once again,  the process starts with the formation of the first layer, during which one site in every two is covered, one-fourth of the sites are occupied twice, and  $5\%$ are occupied three times.

\section{Large $k$ limit: simulation and extrapolation}
\label{sec:sim}

The number of equations increases roughly as $k^2$, which poses significant limitations for obtaining exact results in the large $k$ limit. To address this, we developed a simple simulation code in Python that allows us to obtain accurate numerical results for various values of $k$. For the RSC of a ring of length $L$, the maximum number of $k$-mers is $L-k$, while the minimum is $1+\lfloor L/k\rfloor$. These upper and lower bounds are the consequences of the basic rule of RSC: A $k$-mer can land on the lattice under the condition that at least one empty site is covered. 

The bounds are finite, so a rejection-free algorithm completes the covering process after a finite number of steps. The algorithm operates on a finite lattice with periodic boundary conditions and proceeds as follows. After adding a $k$-mer with the leftmost site $i$, we remove the site $i$ from the list of available sites. The sites on the right, $j\in [i+1,j+k-1]$, are also removed if the site $j+k$ is occupied. Similarly, the sites on the left, $j\in [i-1,i-k]$, are removed if the site $j-k$ is occupied.

\begin{figure}
	\centering
	\includegraphics[width=0.8\columnwidth]{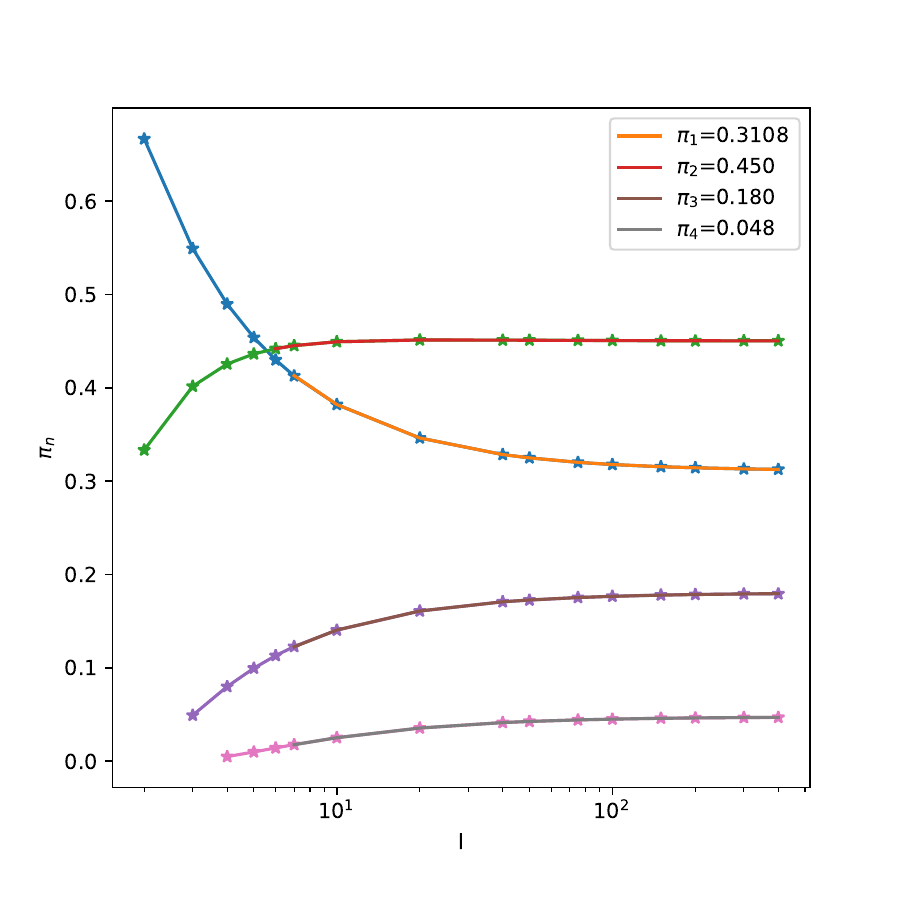}
	\caption{The final (jammed) densities $\pi_1^{(k)}, ~\pi_2^{(k)}, ~\pi_3^{(k)}, ~\pi_4^{(k)}$ versus $k$.}. 
	\label{fig:1}
\end{figure}

We performed $10^6$ independent realizations with ring size ranging from $L=150$ to $1000$ for different values of $k$ ranging from $2$ to $150$. The mean values and standard deviations of site occupations were calculated and are summarized in Table I and displayed in Fig.~\eqref{fig:1}.  
	
	\begin{table}
		\begin{center}
		\begin{tabular}{|c|c|c|c|c|c|}
			\hline
			$k$ &$\pi_1$ &$\pi_2$&$\pi_3$&$\pi_4$& $\pi_5$\\
			\hline\hline
			2& 0.6666&0.33333   &x&x&x \\
			\hline
			3 &0.5492&0.4016 &0.0492  &x &x\\
			\hline
			4 &0.4897& 0.4254& 0.08& 0.00488 &x\\
			\hline
			5 & 0.4538& 0.4363& 0.0996& 0.00988& 0.000364 \\
			\hline
			6& 0.4299& 0.4419& 0.113& 0.0142& 0.000878\\
			\hline
			7& 0.4128& 0.4451& 0.1228& 0.0177& 0.00142 \\
			\hline
			10 &0.3821& 0.4493& 0.1404& 0.0251& 0.00287\\
			\hline
			20 &0.3462& 0.4512& 0.1609& 0.0355& 0.00549 \\
			\hline
			40& 0.3285& 0.4510& 0.1708 &0.04137&0.00728\\
			\hline
			50 &0.3249& 0.4510& 0.1726& 0.0426& 0.00766\\
			\hline
			75 &0.3201& 0.4508 &0.1753 &0.04418&0.00824\\
			\hline
			100& 0.3178& 0.4506& 0.1767& 0.0450& 0.00846 \\
			\hline
			150&0.3154& 0.4503& 0.1780& 0.0460& 0.00876 \\
			\hline
			200&0.3143& 0.4502& 0.17857& 0.0464& 0.00891\\
			\hline
			400 &0.3126 &0.4503 &0.17928& 0.0469&0.009191 \\
			\hline
		\end{tabular}		
			\end{center}
\caption{The jammed densities $\pi_1^{(k)}, ~\pi_2^{(k)}, ~\pi_3^{(k)}, ~\pi_4^{(k)}, ~\pi_5^{(k)}$  for $k=2, \ldots,7$ and several $k$ in the range from $10$ to 400.}
	\end{table}

Simulations reveal that $\pi_1$ is a decreasing function of $k$.  All densities $\pi_j$ with $j\geq 1$ converge to finite values in the $k\to \infty$ limit. Figure \eqref{fig:1} shows the evolution of the final densities $\pi_j^{(k)}(\infty)$ versus $k$ (dots). The full curves correspond to the best fit at large $k$ and the labels display the estimated asymptotic values of the probabilities when $k$ goes to $ \infty$. 

The simplest non-trivial satuation probabilty  $\pi_1^{(k)}$ is a decreasing function of $k$. This assertion is supported by analytical results for $k=2,3,4,5$ and simulation results for $k\leq 400$. The convergence to $\pi_1^{(\infty)}$ is algebraic
\begin{equation}
	\pi_1^{(k)} = \pi_1^{(\infty)} + \frac{C_1}{k}+ \frac{C_2}{k^2}+\ldots
\end{equation}
so it can be improved by performing the Richardson extrapolation. By definition  \cite{Bender}
\begin{equation}
	\label{Richardson}
	R\big[\pi_1^{(k)}\big] = k \pi_1^{(k)} - (k-1) \pi_1^{(k-1)}
\end{equation}

Table~\ref{Table:Richardson} shows exact predictions for the Richardson extrapolation: $R\big[\pi_1^{(2)}\big]=\frac{1}{3}, R\big[\pi_1^{(3)}\big]=\frac{11}{35}, R\big[\pi_1^{(4)}\big]=\frac{14}{45}$ and $R\big[\pi_1^{(5)}\big]=\frac{239}{770  }$. It looks that already for small $k$, the Richardson extrapolation $R\big[\pi_1^{(k)}\big]$ is strikingly close to $\pi_1^{(\infty)}$, e.g., $R\big[\pi_1^{(3)}\big]$ seems closer than $\pi_1^{(200)}$ to the limiting value $\pi_1^{(\infty)}$. Thus, simulation results and Richardson extrapolations suggest
\begin{equation}
	\pi_1^{(\infty)}\approx 0.310
	\end{equation}

\begin{table}[h!]
	\centering
	\renewcommand{\arraystretch}{1.5}{
		\begin{tabular}{| c | l | l | l | l | l |l |}
			\hline
			$k$            & $1$   & $2$                        & \qquad$3$                                    & ~ \qquad$4$  &~ ~\qquad$5$            & ~~$200$\\ 
			\hline
			$\pi_1$      & $1$       & $\frac{2}{3}$       & $\frac{173}{315}\approx 0.5492$       & $\frac{617}{1260}\approx 0.48968$&$\frac{629}{1386}\approx0.453823$   & $0.3143$  \\
			\hline
			$R\big[\pi_1\big]$   & & $\frac{1}{3}$   & $\frac{11}{35}  \approx 0.3142$  & $\frac{14}{45}=0.3111\ldots$  &  $\frac{239}{770}=0.3103896$ &     \\
			\hline
			\hline
			$\pi_2$    &  &    $\frac{1}{3}$&    $\frac{253}{630}\approx 0.40158$          & $\frac{10723}{25200}\approx 0.4255$       & $\frac{294797}{675675}\approx0.4362$   & $ 0.4502$  \\
			\hline
			$R\big[\pi_2\big]$   & &    & $\frac{113}{210}  \approx 0.53809$  & $\frac{3133}{6300}\approx 0.49730$    &  $\frac{1295773}{2702700}\approx 0.4794$   &     \\
			\hline
		\end{tabular}
	}
	\caption{The fractions of single-covered and double-covered sites $\pi_1^{(k)}$ and $\pi_2^{(k)}$: analytical results for $k=1,2,3,4,5$ and numerical result for $k=200$. The Richardson extrapolation defined by Eq.~\eqref{Richardson} is also shown.} 
	\label{Table:Richardson}
\end{table}

Table~\ref{Table:1234} collects analytical predictions for the jammed densities, Eqs.~\eqref{jammed12}, \eqref{jammed:123}, and \eqref{jammed:1234}. 

\begin{table}[h!]
	\centering
	\renewcommand{\arraystretch}{1.5}{
		\begin{tabular}{| c | c | c | c | c |c |}
			\hline
			$k$                  & ~$1$~    & ~$2$~                 & $3$                            & $4$       &$5$                        \\ 
			\hline
			$\pi_1^{(k)}$     & $1$       & $\frac{2}{3}$     & $\frac{173}{315}$      & $\frac{617}{1260}$  &   $\frac{629}{1386}$  \\
			\hline
			$\pi_2^{(k)}$     &              & $\frac{1}{3}$     & ~$\frac{253}{630}$~  & $\frac{10723}{25200}$ &$\frac{294797}{675675}$ \\
			\hline
			$\pi_3^{(k)}$     &              &                          & $\frac{31}{630}$         & ~$\frac{1007}{12600}$~ &$\frac{76931}{772200}$  \\
			\hline
			$~\pi_4^{(k)}$~  &              &                          &                                    & $\frac{41}{8400}$  &$\frac{347}{35100}  $      \\
			\hline
			$~\pi_5^{(k)}$~  &              &                          &                                   & & $\frac{179}{491400}$\\
				\hline
		\end{tabular}
	}
	\caption{Available analytical results for the final (jammed) densities $\pi_j^{(k)}$ with $k=1,2,3, 4,5$ and $j\leq k$. } 
	\label{Table:1234}
\end{table}

\section{Discussion}
\label{sec:discussion}

We demonstrated that the RSC of the one-dimensional lattice by $k$-mers is, in principle, solvable. The analysis becomes very laborious when $k$ increases, and we obtain explicit exact results when $k\leq 5$. Our study reveals that the final (jammed) densities are always rational numbers. The same feature is observed in dynamical space packing \cite{Dandekar2023}. The RSC kinetics exhibits an interesting feature: The fraction $\pi_1^{(k)}(t)$ of single-occupied sites has a (shallow) peak at a certain finite time for all $k\geq 3$; other densities $\pi_j^{(k)}(t)$ with $2\leq j\leq k$ are monotonically increasing functions of time. 

It is worth noting that regardless of $k$, the covering rapidly converges as the densities $\pi_j^{(k)}(\infty)$ are quickly decaying functions of the `height' $j$. However, obtaining a solution in the large $k$ limit remains a challenging problem, as the complexity of the model increases rapidly with $k$. This contrasts with random sequential adsorption models \cite{Renyi58,Evans93,Talbot00}, where the {\em screening property} applies to both lattice and off-lattice models. The screning property means that the evolution of free intervals is independent. The RSC models possess a {\em weak screening property}, wherein larger intervals, whose states can be traced from the past, ultimately close the hierarchy of equations with empty intervals that only satisfy the screening property.

In RSA models, the solution of the kinetic equations\cite{Renyi58,Evans93} has opened the door to obtaining a large number of observables (pair correlation functions \cite{Bonnier1994}, large deviation functions\cite{Krapivsky20}, etc.) that allow for a comprehensive description of the RSA models. It would be interesting to analyze similar observables in the realm of the RSC models. Another direction of the future work is the analysis of  the cluster structure and the entropy of the model similar to work for the RSA models and some other models, see \cite{Luck2024,Krapivsky2023b} and references therein.

\appendix
\section{Derivation for the pentamer model}
\label{sec:pentamerappendix}

For the pentamer model, the closure of the hierarchy of occupation probabilities still holds, but the process is tedious. To provide clarity, we present some details of the derivation here. The joint probabilities $p_{00011}$ and $p_{00012}$ satisfy 
\begin{equation}
	\begin{split}\label{eq:p3p4}
		\frac{d p_{00011} }{dt}&=-7p_{00011}-p_{000111110}+p_{00000}
		\\
		\frac{d p_{00012}}{dt}&=-7p_{00012}+p_{000111110}+p_{00001} 
	\end{split} 
\end{equation}
We know the probabilities $p_{00000}=E_5^{(5)}$ and $p_{00001}=S_4^{(5)}$, see \eqref{Em-3-sol} and \eqref{Sm-sol}, and we also know 
\begin{equation}
	p_{000111110}=e^{-12t}(1-e^{-t})	
\end{equation}
Using these results we solve \eqref{eq:p3p4} to yield
\begin{equation}
	\begin{split}\label{eq:p3p4sol}
		p_{00011} &=e^{-7 t} (1-e^{-t})
		\left(\frac{7}{15}+\frac{7}{15} e^{-t}-\frac{1}{30} e^{-2t}-\frac{1}{30} e^{-3t}-\frac{1}{30} e^{-4t}+\frac{1}{6} e^{-5t}\right)
		\\
		p_{00012}&=
		e^{-7 t} (1-e^{-t})^2\left(\frac{8}{15}+\frac{1}{15} e^{-t}+\frac{1}{10} e^{-2t} +\frac{2}{15} e^{-3t}+\frac{1}{6} e^{-4t}\right)	
	\end{split} 
\end{equation}
The joint probabilities $p_{00111}$ and $p_{00112}$ satisfy 
\begin{equation}
\label{111-112}
	\begin{split}
		\frac{d p_{00111} }{dt}&=-6p_{00111}-2p_{00111110}+p_{00000}
		\\
		\frac{d p_{00112}}{dt}&=-6p_{00112}+p_{00111110}+p_{00011} 
	\end{split} 
\end{equation}
Additional probabilities that appear in Eqs.~\eqref{111-112} are known, e.g.,
\begin{equation}
	p_{00111110}=e^{-11t}(1-e^{-t})	
\end{equation}
We thus solve \eqref{111-112} and find
\begin{equation}\label{eq:p5p6}
	\begin{split}
	p_{00111} &={\mathrm e}^{-6 t} (1-{\mathrm e}^{-t}) \left(\frac{4}{15}+\frac{4}{15} {\mathrm e}^{- t}+\frac{4}{15}{\mathrm e}^{-2 t}-\frac{1}{15} {\mathrm e}^{-3 t}-\frac{1}{15} {\mathrm e}^{-4 t}+\frac{1}{3} {\mathrm e}^{-5 t} \right)
		\\
	p_{00112} &= {\mathrm e}^{-6 t} (1-{\mathrm e}^{-t})^2\left(\frac{1}{5}+\frac{2}{5} {\mathrm e}^{- t}+\frac{1}{10} {\mathrm e}^{- 2t}+\frac{2}{15} {\mathrm e}^{- 3t}+\frac{1}{6} {\mathrm e}^{- 4t}  \right)
	\end{split} 
\end{equation}
Similarly we write the governing equations for two more joint probabilities 
\begin{equation}
	\begin{split}
		\frac{d p_{00122} }{dt}&=-6p_{00122} -p_{001222210} +p_{00111110}+p_{00011}
		\\
		\frac{d p_{00123}}{dt}&=-6p_{00123} +p_{001222110} +p_{00012}
	\end{split} 
\end{equation}
and solve them to give
\begin{equation}
	\begin{split}\label{eq:p7p8}
		p_{00122} &={\mathrm e}^{-6 t} (1-{\mathrm e}^{-t})^2\left(\frac{1}{3}+\frac{1}{5} {\mathrm e}^{- t}+\frac{1}{15} {\mathrm e}^{-2 t}+\frac{1}{10} {\mathrm e}^{-3 t}+\frac{2}{15} {\mathrm e}^{-4 t}+\frac{1}{6} {\mathrm e}^{-5 t} \right)
		\\
		p_{00123}&={\mathrm e}^{-6t} (1-{\mathrm e}^{-t})^3\left( \frac{1}{5}+\frac{1}{15} {\mathrm e}^{-t}+\frac{1}{10} {\mathrm e}^{-2t}+\frac{2}{15} {\mathrm e}^{-7t}+\frac{1}{6} {\mathrm e}^{-4t}\right)
	\end{split} 
\end{equation}
The evolution equations for joint probabilities with a single vacant site
\begin{equation}
\label{1111-1112}
	\begin{split}
		\frac{dp_{01111} }{dt}&=-5p_{01111} -3p_{0111110}+p_{00000}
		\\
		\frac{d p_{01112} }{dt}&=-5p_{01112}  +p_{0111110} +p_{00001}
	\end{split} 
\end{equation}
depend on $p_{0111110}=e^{-10t}(1-e^{-t})$. Using this result we solve \eqref{1111-1112} and find
\begin{equation}\label{eq:p9p10}
	\begin{split} 
		p_{01111} &={\mathrm e}^{-5 t} (1-{\mathrm e}^{-t})\left(\frac{3}{20}+\frac{3}{20} {\mathrm e}^{-t}+\frac{3}{20} {\mathrm e}^{-2t}+\frac{3}{20} {\mathrm e}^{-3t}-\frac{1}{10} {\mathrm e}^{-4t}+\frac{1}{2} {\mathrm e}^{-5t} \right)
		\\
		p_{01112} &={\mathrm e}^{-5 t} (1-{\mathrm e}^{-t})^2\left(\frac{7}{60}+\frac{7}{30} {\mathrm e}^{-t}+\frac{7}{20} {\mathrm e}^{-2 t}+\frac{2}{15} {\mathrm e}^{-3 t}+\frac{1}{6} {\mathrm e}^{-4 t}\right)
	\end{split} 
\end{equation}
Another pair of equations for joint probabilities with a single vacant site
\begin{equation}
\label{1122-1123}
	\begin{split}
		\frac{dp_{01122} }{dt}&=-5p_{01122} -p_{011222110}+p_{00011}+p_{0111110}
		\\
		\frac{dp_{01123}}{dt}&=-5p_{01123}  +p_{011222110}  +p_{00012}
	\end{split}  
\end{equation}
can be also solved since all additional joint probabilities are known. In particular 
\begin{equation}
	p_{011222110}={\mathrm e}^{-10 t}(1-{\mathrm e}^{-t})^2 \left(\frac{1}{3}+\frac{2}{3}{\mathrm e}^{-t}\right) 
\end{equation}
The solution of \eqref{1122-1123} reads 
\begin{equation}
	\begin{split}
		p_{01122} &={\mathrm e}^{-5t}(1-{\mathrm e}^{-t})^2 \left(
		\frac{239}{1680}+\frac{239}{840} {\mathrm e}^{-t} +\frac{65}{336} {\mathrm e}^{-2t}+\frac{43}{420} {\mathrm e}^{-3t}+\frac{229}{1680} {\mathrm e}^{-4t}+\frac{31}{840} {\mathrm e}^{-5t}+\frac{5}{48} {\mathrm e}^{-6t}
		  \right) \\
		p_{01123}&= {\mathrm e}^{-5 t} (1-{\mathrm e}^{-t})^3\left( \frac{97}{1680}+\frac{97}{560} {\mathrm e}^{-t}+\frac{67}{840} {\mathrm e}^{-2 t}+\frac{31}{280} {\mathrm e}^{-3 t}+\frac{79}{560} {\mathrm e}^{-4 t}\right.\\
		&\left.+\frac{5}{48} {\mathrm e}^{-5 t} \right)
	\end{split}  
\end{equation}
One more pair of equations for joint probabilities with a single vacant site
\begin{equation}
	\begin{split}
		\frac{dp_{01222} }{dt}&=-5p_{01222}-2p_{01222210}+p_{00111}+p_{0111110}
		\\
		\frac{dp_{01223} }{dt}&=-5p_{01223} +p_{01222210}+p_{00112}
	\end{split}  
\end{equation}
is solved to yield
\begin{equation}
	\begin{split}
		p_{01222} &={\mathrm e}^{-5 t} ({1-\mathrm e}^{-t})^2\left(\frac{13}{60}+\frac{1}{6} {\mathrm e}^{- t}+\frac{7}{60} {\mathrm e}^{-2 t}+\frac{1}{15} {\mathrm e}^{-3 t}+\frac{1}{10} {\mathrm e}^{-4 t}+\frac{1}{3} {\mathrm e}^{-5 t} \right)  \\
		p_{01223} &={\mathrm e}^{-5 t} (1-{\mathrm e}^{-t})^3\left(\frac{7}{60}+\frac{3}{20} {\mathrm e}^{- t}+\frac{1}{10} {\mathrm e}^{-2 t}+\frac{2}{15} {\mathrm e}^{-3 t}+\frac{1}{6} {\mathrm e}^{-4 t} \right)
	\end{split}  
\end{equation}
To solve another pair of equations for joint probabilities with a single vacant site
\begin{equation}
	\begin{split}
		\frac{dp_{01233} }{dt}&=-5p_{01233}-p_{012333210}+p_{01222210}+p_{00122}
		\\
		\frac{dp_{01234} }{dt} &=-5p_{01234}+p_{00123}+p_{012333210}
	\end{split}  
\end{equation}
we use
\begin{equation}
	p_{012333210} = \frac{2 {\mathrm e}^{-13 t} \left({\mathrm e}^{t}-1\right)^3}{3}
\end{equation}
and find
\begin{equation}
	\begin{split}
		p_{01233} &= {\mathrm e}^{-5 t}(1- {\mathrm e}^{- t})^3 \left(\frac{79}{560}+\frac{151}{1680} {\mathrm e}^{- t}+\frac{67}{840} {\mathrm e}^{-2 t}+\frac{31}{280} {\mathrm e}^{-3 t}+\frac{79}{560} {\mathrm e}^{-4 t}+\frac{5}{48} {\mathrm e}^{-5 t} \right)   \\
		p_{01234} &= {\mathrm e}^{-5 t}(1- {\mathrm e}^{- t})^4 \left( \frac{33}{560}+\frac{1}{28} {\mathrm e}^{- t}+\frac{47}{840} {\mathrm e}^{-2 t}+\frac{11}{140} {\mathrm e}^{-3 t}+\frac{5}{48} {\mathrm e}^{-4 t}\right) 
	\end{split}  
\end{equation}
Similarly we compute
\begin{equation}
	\begin{split}
		p_{10001}&={\mathrm e}^{-7 t} \left(1-{\mathrm e}^{-t}\right)^{2}
		\\
		p_{10011} &={\mathrm e}^{-6 t} \left(1-{\mathrm e}^{-t}\right)^{2}\left(\frac{7}{15}+\frac{7}{15} {\mathrm e}^{-t}-\frac{1}{30} {\mathrm e}^{-2 t}-\frac{1}{30} {\mathrm e}^{-3 t}-\frac{1}{30} {\mathrm e}^{-4 t}+\frac{1}{6} {\mathrm e}^{-5 t}\right)
		\\
		p_{10012} &={\mathrm e}^{-6 t} \left(1-{\mathrm e}^{-t}\right)^{3}\left(\frac{8}{15}+\frac{1}{15} {\mathrm e}^{- t}+\frac{1}{10} {\mathrm e}^{-2 t}+\frac{2}{15} {\mathrm e}^{-3 t}+\frac{1}{6} {\mathrm e}^{-4 t}\right)
		\\
		p_{10111}&= {\mathrm e}^{-5 t} \left(1-{\mathrm e}^{-t}\right)^{2}\left(\frac{4}{15}+\frac{4}{15} {\mathrm e}^{- t}+\frac{4}{15} {\mathrm e}^{-2 t}-\frac{1}{15} {\mathrm e}^{-3 t}-\frac{1}{15} {\mathrm e}^{-4 t}+\frac{1}{3} {\mathrm e}^{-5 t}\right)
		\\
		p_{10112} &= {\mathrm e}^{-5 t} \left(1-{\mathrm e}^{-t}\right)^{3}\left(\frac{1}{5}+\frac{2}{5} {\mathrm e}^{- t}+\frac{1}{10} {\mathrm e}^{-2 t}+\frac{2}{15} {\mathrm e}^{-3 t}+\frac{1}{6} {\mathrm e}^{-4 t}\right)
		\\
		p_{10122} &={\mathrm e}^{-5 t} \left(1-{\mathrm e}^{-t}\right)^{3}\left(\frac{1}{3}+\frac{1}{5} {\mathrm e}^{- t}+\frac{1}{15} {\mathrm e}^{-2 t}+\frac{1}{10} {\mathrm e}^{-3 t}+\frac{2}{15} {\mathrm e}^{-4 t}+\frac{1}{6} {\mathrm e}^{-5 t}\right)
		\\
		p_{10123} &={\mathrm e}^{-5 t} \left(1-{\mathrm e}^{-t}\right)^{4}\left(\frac{1}{5}+\frac{1}{15} {\mathrm e}^{- t}+\frac{1}{10} {\mathrm e}^{-2 t}+\frac{2}{15} {\mathrm e}^{-3 t}+\frac{1}{6} {\mathrm e}^{-6 t}\right)
	\end{split}  
\end{equation}
Other relevant joint probabilities evolve according to 
\begin{equation}
	\begin{split}\label{eq:phigh}
		\frac{dp_{11011}}{dt} &=-5p_{11011}+2p_{00011}-2p_{110111110} \\
		\frac{dp_{110111110}}{dt}&=-10p_{110111110}+p_{110000000}-
		p_{0111110111110}\\
		\frac{dp_{0111110111110}}{dt}&=-15p_{0111110111110}+2p_{0000000111110}\\
		\frac{dp_{0000000111110}}{dt}&=-16p_{0000000111110}+ p_{0000000000000}
	\end{split}  
\end{equation}
Integrating Eq.\eqref{eq:phigh}
\begin{equation}
	\begin{split}
	p_{0000000111110}&=e^{-16t}(1-e^{-t})\\
	p_{0111110111110}&=e^{-15t}(1-e^{-t})^2
	\end{split}  
\end{equation}

One can write the evolution equations for the remaining relevant probabilities $p_{11012}$ and $p_{21012}$, but they can be alternatively determined from the sum rules \eqref{eq:sumrules}
\begin{equation}
	\begin{split}
		p_{11012} &=p_{1101}-p_{11011}\\
		p_{21012}&= p_{1012}-p_{11012}
	\end{split}  
\end{equation}

\section*{References}
%\bibliography{references-packing}

\providecommand{\newblock}{}

\end{document}